\documentclass[10pt, conference]{IEEEtran}
\IEEEoverridecommandlockouts
\usepackage{cite}
\usepackage{amsmath,amssymb,amsfonts}
\usepackage{graphicx}
\usepackage{textcomp}
\usepackage{xcolor}
\usepackage{xurl}
\usepackage{tabularx}
\usepackage{multirow}
\usepackage{xcolor}%
\usepackage{textcomp}%
\usepackage{manyfoot}%
\usepackage{booktabs}%
\usepackage{listings}%
\usepackage{arydshln}
\usepackage{adjustbox}
\usepackage{algorithm}
\usepackage{algpseudocode}
\usepackage{comment}
\usepackage{tabularray}
\usepackage{balance}

\def\BibTeX{{\rm B\kern-.05em{\sc i\kern-.025em b}\kern-.08em
    T\kern-.1667em\lower.7ex\hbox{E}\kern-.125emX}}

\newcommand\blfootnote[1]{%
  \begingroup
  \renewcommand\thefootnote{}\footnote{#1}%
  \addtocounter{footnote}{-1}%
  \endgroup
}

\begin{document}

\title{Comparing Performance and Portability between CUDA and SYCL for Protein Database Search on NVIDIA, AMD, and Intel GPUs}








\author{\IEEEauthorblockN{Manuel Costanzo}
\IEEEauthorblockA{ \textit{III-LIDI, Facultad de Informática,} \\ \textit{UNLP - CIC} \\ La Plata, Argentina \\
0000-0002-6937-3943}
\and
\IEEEauthorblockN{Enzo Rucci}
\IEEEauthorblockA{ \textit{III-LIDI, Facultad de Informática,} \\ \textit{UNLP - CIC} \\ La Plata, Argentina \\
0000-0001-6736-7358}
\and
\IEEEauthorblockN{Carlos Garc\'ia-S\'anchez}
\IEEEauthorblockA{\textit{Dpto. Arquitectura de Computadores y Automática,} \\ 
\textit{Universidad Complutense de Madrid} \\ Madrid, España \\
0000-0002-3470-1097}
\and
\IEEEauthorblockN{Marcelo Naiouf}
\IEEEauthorblockA{ \textit{III-LIDI, Facultad de Informática,} \\ \textit{UNLP - CIC} \\ La Plata, Argentina \\
0000-0001-9127-3212}
\and
\IEEEauthorblockN{Manuel Prieto-Mat\'ias}
\IEEEauthorblockA{\textit{Dpto. Arquitectura de Computadores y Automática,} \\ 
\textit{Universidad Complutense de Madrid} \\ Madrid, España \\
0000-0003-0687-3737}
}

\maketitle

\begin{abstract}
The heterogeneous computing paradigm has led to the need for portable and efficient programming solutions that can leverage the capabilities of various hardware devices, such as NVIDIA, Intel, and AMD GPUs. This study evaluates the portability and performance of the SYCL and CUDA languages for one fundamental bioinformatics application (Smith-Waterman protein database search) across different GPU architectures, considering single and multi-GPU configurations from different vendors. The experimental work showed that, while both CUDA and SYCL versions achieve similar performance on NVIDIA devices, the latter demonstrated remarkable code portability to other GPU architectures, such as AMD and Intel. Furthermore, the architectural efficiency rates achieved on these devices were superior in 3 of the 4 cases tested. This brief study highlights the potential of SYCL as a viable solution for achieving both performance and portability in the heterogeneous computing ecosystem.

\blfootnote{\copyright 2023 IEEE.  Personal use of this material is permitted.  Permission from IEEE must be obtained for all other uses, in any current or future media, including reprinting/republishing this material for advertising or promotional purposes, creating new collective works, for resale or redistribution to servers or lists, or reuse of any copyrighted component of this work in other works.

The final authenticated version is available online at \url{https://doi.org/10.1109/SBAC-PAD59825.2023.00023}}

\end{abstract}


\section{Introduction}
\label{sec:intro}

In the last decade, the quest to improve the energy efficiency of computing systems has fueled the trend toward heterogeneous computing and massively parallel architectures~\cite{Giefers2016}. 
Nowadays, GPUs can be considered the dominant accelerator, and Nvidia, Intel, and AMD are the most prominent manufacturers. 
In the 4th quarter of 2022, Intel and AMD had 9\% of the market, with Nvidia dominating the discrete graphics card market at 82\%. Moreover, considering also the integrated and embedded graphics, Intel had 71\% quote, Nvidia 17\%, and AMD 12\%~\footnote{\url{https://www.pcgamer.com/intel-is-already-matching-amd-for-gaming-graphics-market-share/}}.
This poses a significant challenge for researchers who use GPUs for their experiments and simulations. The critical question is how to use this growing computational capacity transparently without having to pay attention to the programming models, hardware support, or mandatory software ecosystem.

Focusing on the programming aspect, CUDA is still the most popular programming language for GPUs, although it is a proprietary language only valid for NVIDIA devices. Fortunately, other open initiatives have contemplated the programming of GPUs or even other accelerators generically. In particular, SYCL is one of the most promising recent initiatives, which is an open standard from the Khronos Group. One noteworthy feature of SYCL is its status as a cross-platform abstraction layer, enabling programmers to adhere to the fundamental principle of "write code once and run it anywhere". In this sense, the same SYCL code can run not only on multiple vendor GPUs but also on different hardware platforms, including CPUs and FPGAs. SYCL capitalizes on programming productivity by reducing the effort required during development tasks and minimizing maintenance costs. The concept of \textit{performance portability} becomes fundamental in this context. Specifically, performance portability encompasses two key aspects: (1) enabling the execution of a single application on various hardware platforms, and (2) achieving a desired level of performance across these diverse platforms~\cite{performance_portability_paper_rev}.

This paper aims to address the previous issue by exploring the SYCL programming paradigm in the field of Bioinformatics and Computational Biology. These research areas have been leveraging GPUs for over two decades~\cite{GPUsInBioinformatics2016} and numerous of their implementations are based on CUDA, imposing significant limitations on portability across a wide range of heterogeneous architectures. For that reason, this study evaluates the portability and performance of the SYCL and CUDA languages for one fundamental bioinformatics application (Smith-Waterman biological sequence alignment) across different GPU architectures, considering single and multi-GPU configurations from multiple vendors. Hence, we select the \textit{SW\#} suite~\cite{swsharp,swsharpdb}: a CUDA-based, memory-efficient implementation for biological sequence alignment, that has been recently migrated to SYCL~\cite{Costanzo2022IWBBIO}.
   Our main contributions can be summarized as:
\begin{itemize}
    \item An adaptation and extension of the performance model from~\cite{lan2017swhybrid}. This performance model is adapted to the features of the \textit{SW\#} suite and also extended to include AMD and Intel GPUs (both discrete and integrated types).
    \item A functional and performance portability study for  \textit{SW\#} applications across different GPU architectures, considering single and multi-GPU configurations from multiple vendors. To the best of our knowledge, no previous study has considered such a diverse and large set of GPUs.
\end{itemize}

The rest of the paper is organized as follows. Section~\ref{sec:back} introduces the background for this research. Section~\ref{sec:imps} describes the case-study applications and also the adapted and extended performance model. Section~\ref{sec:results} presents the functional and performance portability results. Finally,  Section~\ref{sec:relworks} discusses some related works, and Section~\ref{sec:conc} presents the conclusions and possible lines for future work.

\section{Background}
\label{sec:back}

\subsection{GPUs and Programming Models}
In 2007, Nvidia introduced CUDA~\cite{CUDA_handson}  alongside the Tesla GPU, to enable general-purpose programming on GPUs. CUDA is a programming model and parallel computing platform specifically designed for general computing on GPUs. While CUDA has become the most popular low-level programming model for general-purpose GPU computing, its main limitation is that it only supports NVIDIA devices. In the opposite sense, OpenCL~\cite{OpenCL} gained prominence because it can be used in several devices and vendors requiring a similar abstraction level as CUDA.

High-Level Programming initiatives such as OpenMP~\cite{OpenMP}, OpenACC~\cite{OpenACC, OpenACC_book}, and SYCL~\cite{SYCL} have played significant roles in the field of parallel computing in GPU scenarios. OpenMP initially focused on multi-core CPU computing but later expanded its support to include accelerators like GPUs with the release of v4.0.  
While OpenACC~\cite{WienkePaul12} (Open Accelerators) emerged as one of the earliest high-level approaches for GPU programming through the use of directive-based programming, OpenMP has even started overshadowing it by incorporating most of their features.

Currently, one of the most promising initiatives in the GPU programming ecosystem is SYCL~\cite{SYCL}. It enables developers to write code for heterogeneous processors using standard ISO C++. It incorporates host and kernel code in a single source file and utilizes templates and lambda functions for generic programming. Moreover, SYCL supports various acceleration APIs, such as OpenCL, enabling seamless integration with lower-level code.

Multiple SYCL implementations are available nowadays: Codeplay's ComputeCpp~\cite{ComputeCpp}, oneAPI by Intel ~\cite{DPCPP}, triSYCL~\cite{triSYCL} led by Xilinx, and OpenSYCL~\cite{hipSYCL} (previously denoted as hipSYCL) led by Heidelberg University. In particular, Intel oneAPI can be considered the most mature developer suite. Among the main features of oneAPI, we can find that is an open, cross-industry project that aims to provide an efficient, high-performance programming model. It eliminates the concept of separate code bases for host and device such as in OpenCL. Moreover, multiple programming languages and different tools for each architecture are supported. Data Parallel C++ (DPC++) is oneAPI's core language for programming accelerators and multiprocessors~\cite{DPCPP}, which integrates SYCL and OpenCL standards without additional extensions. Additionally, oneAPI facilitates interoperability with optimized libraries such as oneCCL, oneDAL, oneDNN, oneMKL, oneTBB, and oneVPL, catering to diverse parallel application domains.

\subsection{Smith-Waterman Algorithm}
\label{sec:SW-Algorithm}

The SW algorithm is widely used to obtain the optimal local alignment between two sequences~\cite{Smith1981}. This method is based on a dynamic programming approach and is highly sensitive since it explores all possible alignments between the sequences.

Given two sequences $Q$ and $D$ of length $|Q|=m$ and $|D|=n$, the recurrence relations for the SW algorithm with the modification of Gotoh~\cite{gotoh81} are defined as follows:


\begin{equation}
	H_{i,j}=max
	\begin{cases}
		0\\
		H_{i-1,j-1}+SM(Q[i],D[j])\\
		E_{i,j}\\
		F_{i,j}
	\end{cases}
	\label{eq:sw2}
\end{equation}

\begin{equation}
	E_{i,j}=max
	\begin{cases}
		H_{i,j-1} - G_{o}\\
		E_{i,j-1} - G_{e}
	\end{cases}
	\label{eq:sw3}
\end{equation}
	
\begin{equation}
	F_{i,j}=max
	\begin{cases}
		H_{i-1,j} - G_{o}\\
		F_{i-1,j} - G_{e}
	\end{cases}
	\label{eq:sw4}
\end{equation}

The similarity score $H_{i,j}$ is computed to identify a common subsequence; $H_{i,j}$ contains the score for aligning the prefixes $Q[1..i]$ and $D[1..j]$. Moreover, $E_{i,j}$ and $F_{i,j}$ correspond to the scores of prefix $Q[1..i]$ and $D[1..j]$ aligned to a gap, respectively. \emph{SM} denotes the \emph{scoring matrix} and defines the match/mismatch scores between residues. Last, $G_{o}$ and $G_{e}$ refer to the gap open and gap extension penalties, respectively. 

First of all, $H$, $E$ and $F$ must be initialized with 0 when $i = 0$ or $j = 0$. Then, the recurrences should be calculated with $1 \leq i \leq m$ and $1 \leq j \leq n$. The highest value in the $H$ matrix ($S$) corresponds to the optimal local alignment score between $Q[1..i]$ and $D[1..j]$. If required, the optimal local alignment is finally obtained by following a traceback procedure whose starting point is $S$. From a computational point of view, it is important to highlight the computational dependencies of any $H$ element. Any cell can be calculated only after the values of the upper, left, and upper-left neighbors are known; imposing restrictions on how \emph{H} can be processed.

\begin{figure}[!t]
 \centering
      \includegraphics[width=0.75\columnwidth]{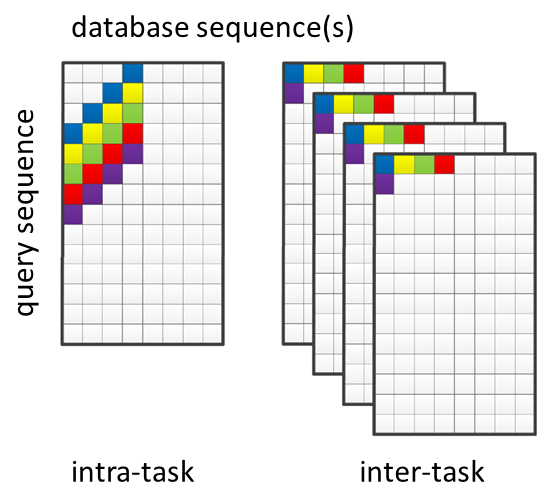}
     \caption{ Parallelization approaches in similarity matrix computations (adapted from~\cite{SWIPE}). Each color indicates the cells that can be computed together in a SIMD manner.}
     \label{fig:schemes}
 \end{figure}

\textbf{SW in practice and parallelization issues.} The SW algorithm can be used to compute: (a) pairwise alignments (one-to-one); usually associated with long DNA sequences; or (b) database similarity searches (one-to-many), usually associated with protein sequence alignment. Although the processing nature of the SW algorithm with the data dependencies on the computation $H_{i,j}$ is very challenging from the point of view of parallelism exploitation, both approaches have been studied in the literature exploiting the SIMD capabilities. In the (a) case, a single matrix is calculated and all Processing Elements (PEs) work collaboratively (\textit{intra-task parallelism}). Due to inherent data dependencies, neighboring PEs communicate to exchange border elements. In the (b) approach, while the intra-task scheme can be used, a better parallel scheme consists in simultaneously calculating multiple matrices without communication between the PEs (\textit{inter-task parallelism})~\cite{SandesReview2016} could be performed. Fig.~\ref{fig:schemes} illustrates both approaches.
 
The SW algorithm runs in quadratic time and space to compute optimal alignment. However, computing optimal alignment scores do not require storing the full similarity matrix and can be calculated in linear space complexity. Similarity database search takes advantage of this feature since optimal alignment only makes sense for very similar sequences. Therefore, all alignment scores are calculated first and optimal alignments are computed only for top-ranked database sequences.

\subsection{Performance portability}

According to Penycook et al.~\cite{performance_portability_paper}, \textit{performance portability} refers to \textit{"A measurement of an application’s performance efficiency for a given problem that can be executed correctly on all platforms in a given set"}. These authors define two different performance efficiency metrics:  \textit{architectural efficiency} and \textit{application
efficiency}. The former denotes the capacity of an application to effectively utilize hardware resources, measured as a proportion of the theoretical peak performance. The latter signifies the application's ability to select the most suitable implementation for each platform, representing a fraction of the highest observed performance achieved.

The metric for performance portability presented by Penycook et al.~\cite{performance_portability_paper} was later reformulated by Marowka~\cite{performance_portability_paper_rev} to address some of its flaws. Formally, for a given set of platforms \textit{H} from the same architecture class, the performance portability $\bar{\Phi}$ of a case-study application $\alpha$ solving problem $p$ is:

$$
\bar{\Phi}(\alpha, p, H)= \begin{cases}\frac{\sum_{i \in H}{e_i(\alpha, p)}}{|H|} & \text { if } i \text { is supported } \forall i \in H \\ \text{not applicable (NA)} & \text { otherwise }\end{cases}
$$

where $e_i(\alpha, p)$ corresponds to the performance efficiency of case-study application $\alpha$ solving problem \textit{p} on the platform \textit{i}.

The \textit{performance portability} concept emphasizes the capability to write code that can efficiently utilize the available computing resources, such as CPUs, GPUs, or specialized accelerators while maintaining high performance regardless of the specific hardware configuration. With performance portability, developers can write code once and have it deliver optimal performance on various target platforms. This eliminates the need for extensive manual code optimizations or platform-specific modifications, reducing development time and effort.

\label{sec:perf-port}

\section{Case-Study Applications and Performance Model}
\label{sec:imps}

\subsection{Case-Study Applications}
\label{sec:case-study}
Two GPU-accelerated implementations of $p=$protein database search were considered for the performance portability evaluation:
\begin{itemize}
    \item \texttt{CUDA}: this version corresponds to the \textit{SW\#} suite, a CUDA-based, memory-efficient implementation for biological sequence alignment, which can be used either as a stand-alone application or a library. It can compute pairwise alignments as well as database similarity searches, for both protein and DNA sequences; and it allows configuring the alignment method (including SW), the open/extension penalties, and the scoring matrix. \textit{SW\#} combines CPU and GPU computation for optimal efficiency. It dynamically balances the workload between the CPU and GPU based on sequence lengths, aiming to minimize idle threads.
    From a parallelization point of view, \textit{SW\#} uses both inter-task and intra-task parallelism but primarily on the GPU side. The GPU divides the workload into two partitions: a ``short kernel'' process shortest database sequences using inter-task scheme, while a ``long kernel'' aligns longest sequences by intra-task strategy. 
    When utilizing multiple GPUs, \textit{SW\#} follows a flexible approach: if the number of query sequences to be aligned is fewer than the number of available GPU devices, all devices align the same query sequence with a different database partition in synchronized manner. Conversely, if the number of query sequences is greater than the number of GPUs, each GPU align a different one against the complete database~\cite{swsharp,swsharpdb}.

    \item \texttt{SYCL}: this code is based on the implementation presented in the paper~\cite{Costanzo2022IWBBIO}, representing a SYCL equivalent. The migration of the \textit{SW\#} suite was performed using \texttt{dpct} (the Data Parallel Compatibility Tool available in the oneAPI suite) and some hand-coding modifications. 
\end{itemize}

\subsection{Performance Model}
\label{sec:perf-model}

Peak theoretical hardware performance must be estimated for all selected GPUs in this study to compute the performance portability metric. This step requires considering both hardware and algorithm features. Fortunately, the previous work from Lan et al.~\cite{lan2017swhybrid} can be used as a basis for this task; in this paper, the computing capability of different devices (including accelerators based on NVIDIA GPUs, Intel CPUs, and the discontinued Intel Xeon Phis) can be estimated using Eq.~\ref{eq:cc-gpu}:

\begin{equation}
\label{eq:cc-gpu}
    Capability = Clock\_Rate \times Throughput \times Lanes
\end{equation}

where \textit{Clock\_rate} refers to the clock frequency, \textit{Throughput} refers to the instruction count that the device can execute in one clock cycle, and \textit{Lanes} refers to the number of SIMD vector lanes. Then, the number of instructions issued in each cell update of the similarity matrix should be counted. In the sequence alignment context, the most popular metric for measuring performance is related to the number of Cell Updated Per Second (CUPS). So the theoretical peak performance of any device could be modeled using Eq.~\ref{eq:theo-peak}:

\begin{equation}
\label{eq:theo-peak}
    Theo\_peak = \frac{Capability}{Instruction\_count\_one\_cell\_update}
\end{equation}

Even though this study only considers GPUs, these equations can serve as a basis to estimate their theoretical peak performance for other devices such as CPUs or FPGAs. For this work, the previous performance model from~\cite{lan2017swhybrid} is adapted to the features of the \textit{SW\#} algorithm and also extended to other GPUs vendors such as AMD and Intel GPUs (both discrete and integrated types). Table~\ref{tab:GCUPS-peak} summarized the theoretical peak performance of selected GPUs using  the Eq.~\ref{eq:theo-peak}. More details can be found in the rest of this section.

\subsubsection{SW\# core instructions}

\textit{SW\#} computes the similarity matrix using 32-bit integers and performs 12 instructions per cell update. Algorithm~\ref{alg:core} presents the snippet of cell update in similarity matrix as in Eq.~\ref{eq:sw2}, Eq.~\ref{eq:sw3}, and Eq.~\ref{eq:sw4}. Just adding, subtracting, and maximum instructions are required to perform a single-cell update.

\begin{algorithm}

\caption{ Core instructions per cell update in similarity matrix}
\label{alg:core}
\begin{algorithmic}[1]
\State $E^{1} = E_{l} - G_{e}$
\Comment{$E_{l}$: E of its left neighbor}
\State $E^{2} = H_{l} - G_{o}$
\Comment{$H_{l}$: H of its left neighbor}
\State $E = max(E^{1},E^{2})$
\State $F^{1} = F_{u} - G_{e}$
\Comment{$F_{u}$: F of its upper neighbor}
\State $F^{2} = H_{u} - G_{o}$
\Comment{$H_{u}$: H of its upper neighbor}
\State $F = max(F^{1},F^{2})$
\State $H = H_{ul} + SM$ \Comment{$H_{ul}$: H of its upper-left neighbor}
\State $H = max(H,E)$
\State $H = max(H,F)$
\State $H = max(H,0)$
\State $A = H$ \Comment $A$: an auxiliary variable
\State $S = max(H,S)$ \Comment $S$: optimal score
\end{algorithmic}
\end{algorithm}

\subsubsection{Architectural features on NVIDIA's GPU}

The \# Cores in an NVIDIA GPU refers to the  number of Streaming Multiprocessors. CUDA does not strictly follow a SIMD execution model but it adopts a similar one denoted as the SIMT model. A \textit{warp} is composed of a group of 32 threads that execute the same instruction stream. According to~\cite{lan2017swhybrid}, "a \textit{warp} in SIMT is equivalent to a \textit{vector} in SIMD, and a \textit{thread} in SIMT is equivalent to a \textit{vector lane} in SIMD". The instruction throughput depends on the CUDA Compute Capability (CC) of each NVIDIA GPU~\footnote{\url{https://docs.nvidia.com/cuda/cuda-c-programming-guide/#maximize-instruction-throughput}}.

\subsubsection{Architectural features on AMD's GPU}

In the RDNA2.0 architecture, the \# Cores represent the number of Compute Units (CUs), which are grouped in pairs into Workgroup Processors (WP). On its behalf, AMD calls \textit{wavefront} and \textit{work-item}  the equivalent of NVIDIA's \textit{warp} and \textit{thread}, respectively. RDNA2.0 supports both wavefront sizes of 32 and 64 work items but the former is prioritized. Each CU contains two SIMD32 vector units, being able to compute 64 add/subtract/max instructions per cycle (\texttt{Int32}). This means that the instruction throughput is 2 for each work item.

\subsubsection{Architectural features on Intel's GPU}

On the discrete segment (dGPUs), Intel has a quite different GPU design philosophy than NVIDIA and AMD. The fundamental block of the Intel Xe microarchitecture is the Xe Core, each of which has 16 Xe Vector Engines (XVEs)~\footnote{Also known as Executions Units (EUs)} that can execute 8 add/subtract/max instructions per cycle  (\texttt{Int32}). Thus, Xe Cores and XVEs map to \# Cores and \# Lanes, respectively, in the proposed model.

On the integrated segment (iGPUs), both Gen9 and Gen12 microarchitectures are similar from a design perspective, differing mainly in the amount of computational resources. In these microarchitectures, the fundamental block is the Subslice, each of which has 8 Execution Units (EUs) that can execute 8 add/subtract/max instructions per cycle  (\texttt{Int32}). Thus, Subslices and EUs refer to \# Cores and \# Lanes, respectively, in the proposed model.

\begin{table*}[t!]
\caption{GPU specifications and their theoretical peak performance in terms of GCUPS}
\label{tab:GCUPS-peak}
\begin{tabular}{|l|cccccc|ccc|c|}
\hline
\textbf{Vendor} &
  \multicolumn{6}{c|}{NVIDIA} &
  \multicolumn{3}{c|}{Intel} &
  AMD \\ \hline
\textbf{Model} &
  \multicolumn{1}{c|}{GTX 980} &
  \multicolumn{1}{c|}{GTX 1080} &
  \multicolumn{1}{c|}{RTX 2070} &
  \multicolumn{1}{c|}{V100} &
  \multicolumn{1}{c|}{RTX 3070} &
  RTX 3090 &
  \multicolumn{1}{c|}{Arc A770} &
  \multicolumn{1}{c|}{UHD 630} &
  UHD770 &
  RX 6700 XT \\ \hline
\textbf{Type} &
  \multicolumn{6}{c|}{Discrete} &
  \multicolumn{1}{c|}{Discrete} &
  \multicolumn{2}{c|}{Integrated} &
  Discrete \\ \hline
\textbf{Microarchitecture} &
  \multicolumn{1}{c|}{\begin{tabular}[c]{@{}c@{}}Maxwell \\ (CC 5.2)\end{tabular}} &
  \multicolumn{1}{c|}{\begin{tabular}[c]{@{}c@{}}Pascal \\ (CC 6.1)\end{tabular}} &
  \multicolumn{1}{c|}{\begin{tabular}[c]{@{}c@{}}Turing \\ (CC 7.5)\end{tabular}} &
  \multicolumn{1}{c|}{\begin{tabular}[c]{@{}c@{}}Volta \\ (CC 7.0)\end{tabular}} &
  \multicolumn{1}{c|}{\begin{tabular}[c]{@{}c@{}}Ampere \\ (CC 8.6)\end{tabular}} &
  \begin{tabular}[c]{@{}c@{}}Ampere \\ (CC 8.6)\end{tabular} &
  \multicolumn{1}{c|}{Xe HPG} &
  \multicolumn{1}{c|}{Gen 9.5} &
  Gen 12.2 &
  RDNA 2.0 \\ \hline
\textbf{\# Cores} &
  \multicolumn{1}{c|}{16} &
  \multicolumn{1}{c|}{20} &
  \multicolumn{1}{c|}{36} &
  \multicolumn{1}{c|}{80} &
  \multicolumn{1}{c|}{46} &
  82 &
  \multicolumn{1}{c|}{32} &
  \multicolumn{1}{c|}{3} &
  4 &
  40 \\ \hline
\textbf{\# Lanes} &
  \multicolumn{1}{c|}{32} &
  \multicolumn{1}{c|}{32} &
  \multicolumn{1}{c|}{32} &
  \multicolumn{1}{c|}{32} &
  \multicolumn{1}{c|}{32} &
  32 &
  \multicolumn{1}{c|}{16} &
  \multicolumn{1}{c|}{8} &
  8 &
  32 \\ \hline
\textbf{\begin{tabular}[c]{@{}c@{}}Instruction \\ throughput\end{tabular}} &
  \multicolumn{1}{c|}{4/2} &
  \multicolumn{1}{c|}{4/2} &
  \multicolumn{1}{c|}{2} &
  \multicolumn{1}{c|}{2} &
  \multicolumn{1}{c|}{2} &
  2 &
  \multicolumn{1}{c|}{8} &
  \multicolumn{1}{c|}{8} &
  8 &
  2 \\ \hline
\textbf{Clock (MHz)} &
  \multicolumn{1}{c|}{1216} &
  \multicolumn{1}{c|}{1733} &
  \multicolumn{1}{c|}{1620} &
  \multicolumn{1}{c|}{1380} &
  \multicolumn{1}{c|}{1725} &
  1695 &
  \multicolumn{1}{c|}{2400} &
  \multicolumn{1}{c|}{1200} &
  1650 &
  2581 \\ \hline
\textbf{\begin{tabular}[c]{@{}l@{}}Theoretical peak \\ (GCUPS)\end{tabular}} &
  \multicolumn{1}{c|}{155.648} &
  \multicolumn{1}{c|}{277.28} &
  \multicolumn{1}{c|}{311.04} &
  \multicolumn{1}{c|}{588.8} &
  \multicolumn{1}{c|}{423.2} &
  741.2 &
  \multicolumn{1}{c|}{819.2} &
  \multicolumn{1}{c|}{19.2} &
  35.2 &
  550.61 \\ \hline
\multicolumn{10}{l}{\begin{tabular}[c]{@{}l@{}}The instruction throughput for GTX 980 and GTX 1080 is 4 for add/subtract and 2 for max/min. The core instructions include 5 \\ add/subtract and 6 max. Thus the equivalent throughput is 3.\end{tabular}} \\
\multicolumn{10}{l}{The core instruction count for each cell update is 12.}  
\end{tabular}
\end{table*}
\section{Experimental Results}
\label{sec:results}

\subsection{Experimental Design}
The experiments were carried out on a set of 10 GPUs, including 6 NVIDIA dGPUs, 1 AMD dGPU, 2  Intel iGPUs, and 1 Intel dGPU. The specific details of these GPUs can be found in Table~\ref{tab:GCUPS-peak}. The oneAPI and CUDA versions used were \texttt{2022.1.0} and \texttt{11.7}, respectively. For both CUDA and SYCL, the optimization flag \texttt{-O3} was used during compilation. To run SYCL code on NVIDIA and AMD GPUs, several modifications had to be made to the build process, as SYCL is not supported by default on these platforms\footnote{\url{https://intel.github.io/llvm-docs/GetStartedGuide.html}} but Codeplay recently has announced free binary plugins\footnote{\url{https://codeplay.com/portal/blogs/2022/12/16/bringing-nvidia-and-amd-support-to-oneapi.html}} to support it. After these modifications, it was possible to run DPC++ code on an NVIDIA GPU using the Clang++ compiler (\texttt{16.0}).

\textit{SW\#} was configured with \texttt{BLOSUM62} as substitution matrix, and 10/2 as insertion/extension gap penalty.  
The flag \texttt{T=0} was also used to remove the impact of the CPU on the final performance (all sequence alignments are computed thoroughly on the GPU).

The performance evaluation was carried out by
searching 20 query protein sequences against the well-known Environmental Non-Redundant database (Env. NR) (\texttt{2021\_04 Release}), which contains 995210546 amino acid residues in 4789355 sequences, with a maximum length of 16925. Query sequences were selected from the Swiss-Prot database~\footnote{Swiss-Prot: ~\url{https://www.uniprot.org/downloads}}, with lengths ranging from 144 to 5478. The access numbers for these queries are: P02232, P05013, P14942, P07327, P01008, P03435, P42357, P21177, Q38941, P27895, P07756, P04775, P19096, P28167, P0C6B8, P20930, P08519, Q7TMA5, P33450, and Q9UKN1. 

In order to minimize fluctuations, the tests were executed 20 times for each set, and the performance was determined based on the average of these multiple runs.


\subsection{Single-GPU Performance and Portability Results}

\begin{figure}[!t]
 \centering
      \includegraphics[width=1\columnwidth]{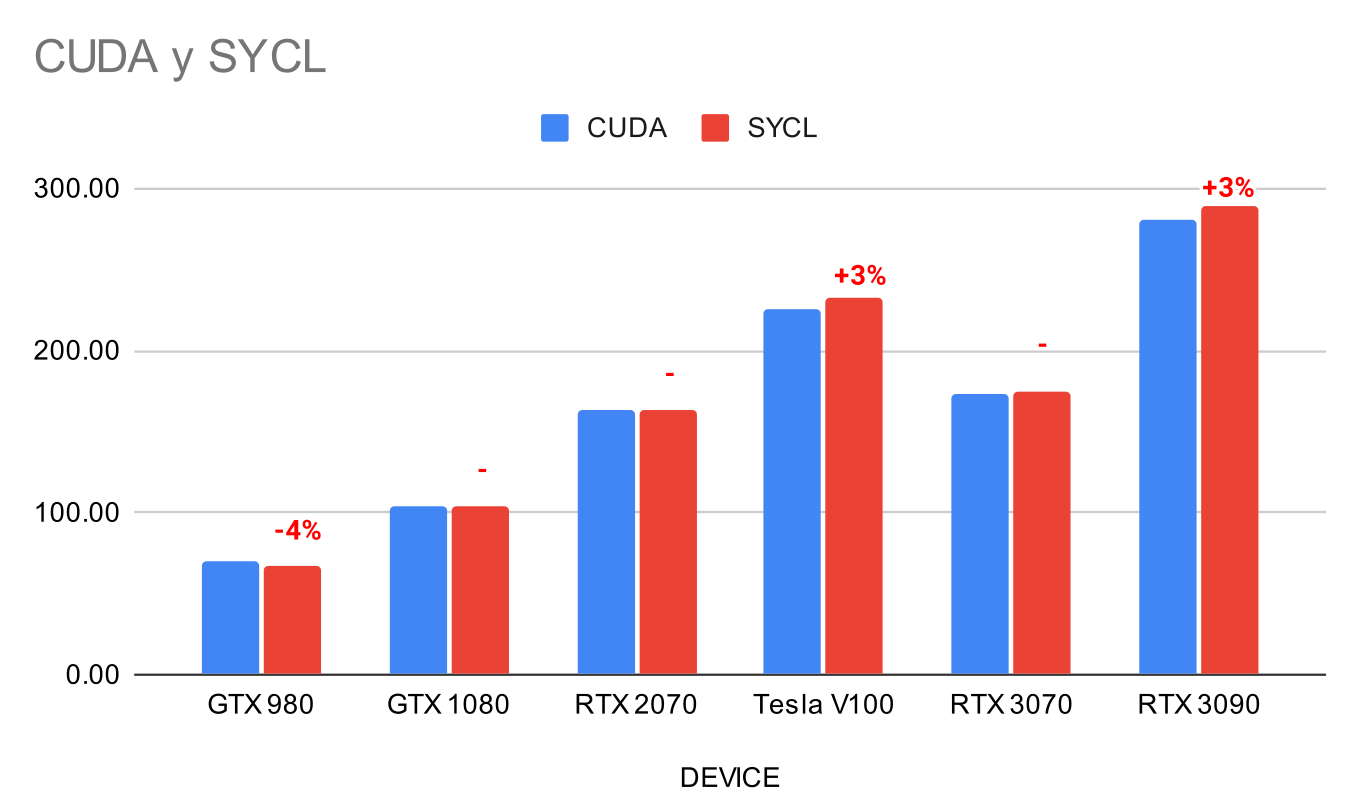}
     \caption{Performance comparison between CUDA and SYCL on single, NVIDIA GPUs}
     \label{fig:nvidia-GPUS}
 \end{figure}

A primary comparison was conducted between the performance of CUDA and SYCL on NVIDIA GPUs (see Fig~\ref{fig:nvidia-GPUS}). As can be seen, both programming models achieve practically the same GCUPS values. On the one hand, the largest performance difference in favor of SYCL was observed on the Tesla V100 (3.4\%). On the other hand, the CUDA implementation did its part on the GTX 980, outperforming SYCL by 4.6\%. Thus, both  CUDA and SYCL are capable of delivering comparable performance for this case study on NVIDIA GPUs.




\begin{table}
\centering
\caption{GCUPS and architectural efficiencies of both CUDA and SYCL codes on single GPUs.}
\label{tab:arch-eff}
\begin{tblr}{
  cells = {c},
  cell{1}{1} = {c=3}{},
  cell{1}{4} = {c=2}{},
  cell{1}{6} = {c=2}{},
  cell{3}{1} = {r=6}{},
  cell{9}{1} = {r=3}{},
  vlines,
  hline{1,9,12-13} = {-}{},
  hline{4-8,10-11} = {2-7}{},
}
\textbf{Platform} &              &                                 & \textbf{CUDA}                                &                                            & \textbf{SYCL}                                &                                        \\
\rotatebox[origin=c]{90}{\textbf{Vendor}}     & \textbf{GPU} & {\textbf{GCUPS}\\\textbf{peak}} & {\textbf{GCUPS}\\\textbf{\textbf{ach.}}} & {\textbf{Arch}\\\textbf{\textbf{eff.}}} & {\textbf{GCUPS}\\\textbf{\textbf{ach.}}} & {\textbf{Arch}\\\textbf{\textbf{eff.}}} \\
\hline

\rotatebox[origin=c]{90}{\textbf{NVIDIA}}            & GTX 980      & 155.5                           & 70.6                                         & 45.3\%                                       & 67.7                                         & 43.5\%                                 \\
                  & GTX 1080     & 277.2                           & 104.5                                        & 37.7\%                                       & 103.8                                        & 37.4\%                                 \\
                  & RTX 2070     & 311.0                           & 162.5                                        & 52.2\%                                       & 163.1                                        & 52.4\%                                 \\
                  & Tesla V100   & 588.8                           & 224.9                                        & 38.2\%                                       & 233.0                                        & 39.5\%                                 \\
                  & RTX 3070     & 423.2                           & 173.1                                        & 40.9\%                                       & 174.4                                        & 41.2\%                                 \\
                  & RTX 3090     & 741.3                           & 280.2                                       & 37.8\%                                       & 288.6                                        & 38.9\%                                 \\

\rotatebox[origin=c]{90}{\textbf{Intel}}             & Arc A770     & 819.2                           & $\times$                                     & NA                                         & 191.4                                        & 23.3\%                                 \\
                  & UHD 630      & 19.2                            & $\times$                                     & NA                                         & 13.1                                         & 68.4\%                                 \\
                  & UHD 770      & 35.2                             & $\times$                                     & NA                                         & 26.6                                         & 75.7\%                                 \\
\rotatebox[origin=c]{90}{\textbf{AMD}}                 & \begin{tabular}[c]{@{}c@{}}RX 6700\\ XT \end{tabular}   & 550.6                           & $\times$                                     & NA  & 284.4                                        & 51.7\%                            
\end{tblr}
\end{table}

Table~\ref{tab:arch-eff} presents a more detailed comparison of the performance and architectural efficiency of CUDA and SYCL codes on NVIDIA, AMD, and Intel GPUs. For each platform, this table shows the peak theoretical performance, the achieved performance for both CUDA and SYCL, and the corresponding architectural efficiency.

On NVIDIA GPUs, CUDA and SYCL demonstrated comparable performance and efficiency values, as was already noted in the analysis from Fig.~\ref{fig:nvidia-GPUS}. As expected, more powerful GPUs are able to achieve higher GCUPS values. As for the architectural efficiency values, they are in the range of 37\%-52\%. It is important to note that, although the highest GCUPS value is presented by RTX 3090 GPU, the most efficient one turns out to be RTX 2070 GPU.

For AMD and Intel GPUs, only the results for SYCL are shown, due to CUDA just supports NVIDIA GPUs. This fact highlights the already mentioned greater portability of SYCL over CUDA. 
It can be said that the results of the SYCL version on these GPUs are generally good. On the one hand, SYCL matches its best efficiency rate on NVIDIA GPUs when running on AMD GPUs. On the other hand, SYCL beats that mark on the 2 integrated GPUs, achieving up to +23.1\% architectural efficiency. The only negative aspect is SYCL's performance on Intel's Arc A770, where performance drops to 23.3\% of architectural efficiency. This value represents its lowest performance and the cause could be related to Intel's discrete GPU design philosophy, which differs from NVIDIA and AMD. However, we plan to profile the code to learn more about this issue.






\begin{table}[htbp]
\caption{Performance portability of both CUDA and SYCL codes on  single GPUs.}
\label{tab:performance-portability}
\begin{center}
\begin{tabular}{|c|c|c|}
\hline
\multicolumn{1}{|c|}{} & \multicolumn{2}{|c|}{$\bar{\Phi}(\alpha,p,H)$} \\

\textbf{Platform set (\textit{H})} & \textbf{CUDA} & \textbf{SYCL} \\

NVIDIA  &	42\% &	42.2\% \\
AMD  &	NA	& 51.7\% \\
Intel (discrete)  &	NA	& 23.3\% \\ 
Intel (integrated)  &	NA	& 72.0\% \\ 
Intel (all) &	NA	& 55.8\% \\ 
\hdashline
NVIDIA $ \cup $ AMD &	NA &	44.3\% \\
NVIDIA $ \cup $ Intel &	NA &	47.2\% \\
Intel $ \cup $ AMD &	NA &	54.8\% \\
\hdashline
NVIDIA $ \cup $ AMD $ \cup $ Intel &	NA &	47.2\% \\
\bottomrule
\end{tabular}
\end{center}
\end{table}

The performance portability of both CUDA and SYCL codes is evaluated in Table~\ref{tab:performance-portability}, where it can be noted that aggregated results are consistent with those observed on an individual basis before. For NVIDIA GPUs, the performance portability of both is quite similar, with values of 42\% and 42.2\%, respectively. As seen before, this indicates that both programming models can deliver a consistent level of performance across the different NVIDIA GPUs used in the tests.

In the case of Intel GPUs, SYCL demonstrated very good architectural efficiency values on the iGPUs, in contrast to the lower efficiency exhibited on the dGPU. Moreover, when considering the combination of AMD and Intel GPUs, SYCL achieves the highest performance portability of the middle set. However, the performance portability decreases when NVIDIA GPUs are also included (last set), as SYCL performance is lower on these devices.


Building on the previous analysis, SYCL consistently outperforms CUDA in terms of performance portability in this study. To be more precise, SYCL achieved nearly the same architectural efficiency as CUDA considering 6 NVIDIA GPUs with 5 different microarchitectures. Moreover, SYCL was not only able to run on multiple vendor GPUs (AMD and Intel), but its architectural efficiency was superior in 3 of the 4 cases tested. This demonstrates not only SYCL's broad compatibility but also its capability to improve performance across a diverse range of GPUs for this application.




\subsection{Multi-GPU Performance and Portability Results}

\begin{figure}[!t]
 \centering
      \includegraphics[width=1\columnwidth]{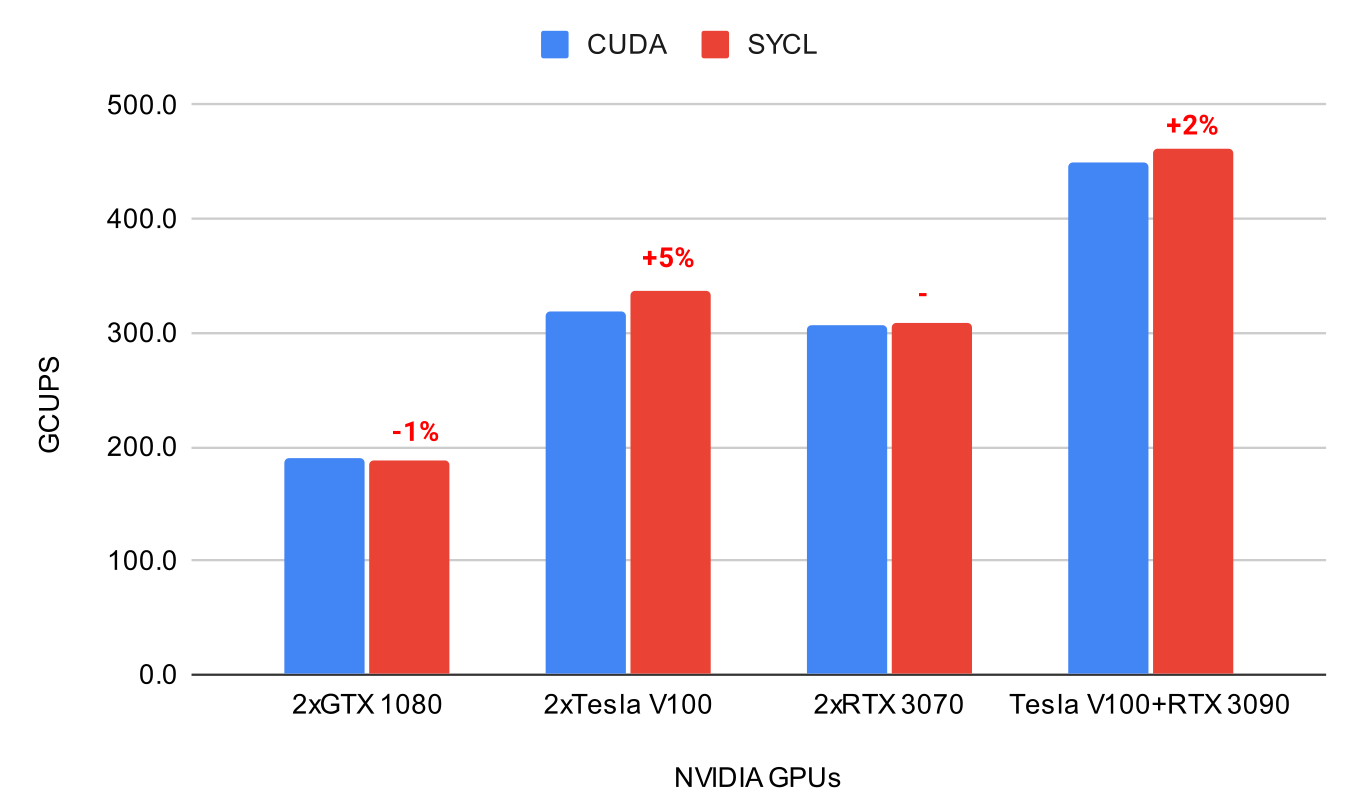}
     \caption{Performance comparison between CUDA and SYCL on multiple NVIDIA GPUs}
     \label{fig:multi-GPUS}
 \end{figure}

To complement the previous single-GPU analysis, a performance comparison was carried out between CUDA and SYCL using different multiple NVIDIA GPUs (see Fig.~\ref{fig:multi-GPUS}). As is the single-GPU case, the two programming models achieve practically the same GCUPS values when NVIDIA devices are used, for both homogeneous and heterogeneous multi-GPU configurations. While CUDA outperforms SYCL when using 2$\times$GTX1080 by approximately 1\%, SYCL achieves the best performance in all other cases, achieving up to 5\% higher GCUPS. Therefore, it can be noted that SYCL does not imply additional overhead when multiple GPUs are used.

Table~\ref{tab:arch-eff-multiple} presents a more detailed comparison of the performance and architectural efficiency of CUDA and SYCL codes on 5 different multi-GPU configurations. It can be seen that for NVIDIA multi-GPUs, the efficiency rates achieved when using 2 GPUs combined are a bit lower than when using a single GPU. This behavior occurs in 3 of the 4 configurations tested (the exception is when using 2$\times$Tesla V100) and can be explained by 2 reasons. On the one hand, it is usual that the efficiency decreases when fixing the problem size and increasing the amount of computational resources. On the other hand, the workload distribution strategy of \textit{SW\#} is very simple, since it distributes the query sequences among the GPUs and does not consider each GPU computing power. Because these sequences do not have the same length, load imbalance can occur between GPUs, reducing performance.

Finally, SYCL once again demonstrates its increased functional portability with Intel's multi-GPU case.  While the performance is not good for the aforementioned reasons, it is interesting to note how SYCL allows using 2 Intel GPUs of different types at the same time: an iGPU and a dGPU.

\begin{table}
\centering
\caption{GCUPS and architectural efficiencies of both CUDA and SYCL codes on multiple GPUs.}
\label{tab:arch-eff-multiple}
\begin{tblr}{
  cells = {c},
  cell{1}{1} = {c=3}{},
  cell{1}{4} = {c=2}{},
  cell{1}{6} = {c=2}{},
  cell{3}{1} = {r=4}{},
  cell{7}{1} = {r=1}{},
  vlines,
  hline{1,7,8} = {-}{},
  hline{4-6} = {2-7}{},
}
\textbf{Platform} &              &                                 & \textbf{CUDA}                                &                                            & \textbf{SYCL}                                &                                        \\
\rotatebox[origin=c]{90}{\textbf{Vendor}}     & \textbf{GPUs} & {\textbf{GCUPS}\\\textbf{peak}} & {\textbf{GCUPS}\\\textbf{\textbf{ach.}}} & {\textbf{Arch}\\\textbf{\textbf{eff.}}} & {\textbf{GCUPS}\\\textbf{\textbf{ach.}}} & {\textbf{Arch}\\\textbf{\textbf{eff.}}} \\
\hline

\rotatebox[origin=c]{90}{\textbf{NVIDIA}} & \begin{tabular}[c]{@{}c@{}}$2\times$ \\ GTX 1080\end{tabular} & 554.6 & 189.8 & 34.2\% & 187.8 & 33.9\% \\
                  & \begin{tabular}[c]{@{}c@{}}$2\times$ \\ Tesla V100 \end{tabular}& 846.4 & 318.1 & 27.0\% & 336.5 & 28.6\% \\
                  & \begin{tabular}[c]{@{}c@{}}$2\times$ \\ RTX 3070 \end{tabular} & 1177.6 & 306.5 & 36.2\% & 308.9 & 36.5\% \\
                  & \begin{tabular}[c]{@{}c@{}} Tesla V100 \\ +\\ RTX 3090\end{tabular} & 1330.1 & 450.5 & 33.8\% & 460.7 & 34.6\% \\

\rotatebox[origin=c]{90}{\textbf{Intel}} & \begin{tabular}[c]{@{}c@{}} Arc A770 \\ +  \\ UHD 770 \end{tabular} & 854.4 & $\times$ & NA & 126.8 & 14.8\% \\
\end{tblr}
\end{table}

\section{Related Works}
\label{sec:relworks}

Some preliminary studies have compared the performance between SYCL and CUDA in different domains. 
In~\cite{mahu20}, the authors employed ADEPT, a GPU-accelerated short-read alignment kernel, as a case study. They found that the SYCL implementation runs approximately $2\times$ slower than its CUDA counterpart in all experiments when using an NVIDIA V100 GPU.  The authors attribute this discrepancy to CUDA's superior utilization of memory cache and SYCL's greater reliance on register usage. Additionally, the authors verified SYCL's code portability on an Intel P630 GPU.

In~\cite{zhe22}, the authors delve into the process of migrating a CPU+GPU application for epistasis detection from CUDA to SYCL, founding that the highest performance of both versions is comparable on an NVIDIA V100 GPU. However, it is important to remark that some hand-tuning was required in the SYCL implementation to reach its maximum performance. When investigating the PTX code, the authors noted that SYCL does not perform the same optimizations as CUDA, such as loop unrolling.

In~\cite{zhem22}, the authors identified performance gaps in several bioinformatics applications. The study involved the selection of open-source applications that had been migrated from CUDA to SYCL, followed by a comprehensive evaluation of their performance on an NVIDIA V100 GPU. Through profiling analysis, the authors found that the SYCL compiler lacks certain optimizations that the CUDA version does, including memory management, instruction vectorization, and loop unrolling, among others.


In~\cite{muh22}, a performance comparison is carried out between SYCL and CUDA in the context of AI models. The authors extend the SYCL-DNN library to include support for NVIDIA GPUs using DPC++ and evaluate its performance against the optimized cuDNN library. Initially, they observed that the non-optimized SYCL-DNN is approximately 50\% slower than cuDNN due to a poorly optimized implementation of SYCL for local memory. However, after using SYCL-BLAS, a significant speedup of up to 90\% of cuDNN's performance is achieved. The remaining 10\% difference is attributed to hand-written, optimized implementations in CUDA.

In~\cite{leo23}, the authors compare two CUDA and SYCL versions of the AutoDock-GPU molecular docking application on an Intel Xeon Platinum 8360Y CPU, an NVIDIA A100 GPU, and an Intel Max 1550 GPU. On the A100 GPU, SYCL exhibits slower performance compared to CUDA in some cases, with performance ratios ranging from 1.24$\times$ to 2.38$\times$. However, in the small test cases, SYCL outperforms CUDA by 1.09$\times$. The authors attribute the lower ratios to the synchronization effort required in compute-intensive regions like the scoring function and gradient calculation. They highlight the need for deeper performance analysis and suggest further optimization, particularly in compute-intensive areas, to improve SYCL performance.

In~\cite{bria20}, the authors analyze the performance of mini-apps that have been created in both SYCL and CUDA, running on an NVIDIA V100 GPU. Even though there are some features not fully supported, SYCL performance is comparable to that of CUDA. Moreover, the performance differences largely stem from variations in memory access patterns.

In~\cite{faquir23}, the author evaluate the gap between performance and code portability in HPC accelerators using the well-known k-means algorithm comparing SYCL with CUDA and OpenMP. SYCL implementation reports higher performance on Intel GPUs and CPUs, equivalent performance on NVIDIA GPUs, and offers potential multi-vendor compatibility.


Unlike the previous works and beyond the results obtained, this performance portability study has considered different GPU architectures, including single and multi-GPU configurations from multiple vendors. To the best of our knowledge, no study has considered such a diverse and large set of GPUs.

\section{Conclusions and Future Work}
\label{sec:conc}

In the field of heterogeneous computing, ensuring functional portability is not trivial for a programming language, and thus providing performance portability represents an even greater challenge. In this study, we address this issue by assessing the portability and performance of the SYCL and CUDA languages for the Smith-Waterman protein database search across
different GPU architectures from multiple vendors. The experimental results show that CUDA and SYCL are capable of delivering comparable performance for this case study on NVIDIA GPUs, including single and multi-GPU configurations. When moving to AMD and Intel GPUs, SYCL was not only able to run on these devices, but its architectural efficiency was superior in 3 of the 4 cases tested. This demonstrates not only SYCL's broad compatibility but also its capability to improve performance across a diverse range of GPUs for this application.

Since SYCL is still an immature programming model,  the positive results found here cannot be generalized; performance will largely depend on the characteristics of the application and the capabilities of the compilers. However, they are a sample of the promising opportunities that SYCL can offer for heterogeneous computing.

Future work will focus on:
\begin{itemize}
    \item Optimizing the SYCL code to reach its maximum performance. In particular, the original \textit{SW\#} suite does not consider some known optimizations for SW alignment~\cite{SWIPE}, such as instruction reordering to reduce their count and the use of lower precision integers to increase parallelism~\footnote{It is important to note that at the time of \textit{SW\#}'s development, most CUDA-enabled GPUs did not support efficient arithmetic on 8-bit vector data types.}. Additionally, improving the workload distribution strategy when using more than one GPU. These improvements will lead to higher efficiency rates.
    \item  Running the SYCL code on other architectures (such as CPUs and CPUs+GPUs) and also considering other SYCL implementations (such as OpenSYCL and ComputeCPP), as well as other programming models like Kokkos~\footnote{\url{https://github.com/kokkos/kokkos}} and RAJA~\footnote{\url{https://github.com/LLNL/RAJA}}, to strengthen the current performance portability study.

\end{itemize}

\balance

\bibliographystyle{IEEEtran}
\bibliography{references}

\end{document}